%% file: main.tex
\documentclass[10pt,comsoc]{IEEEtran} 
\input{input.tex}

\usepackage{xcolor}
\usepackage{tikz}
\usetikzlibrary{arrows.meta, positioning, decorations.pathreplacing, matrix, calc}

\usepackage{pifont, booktabs, array}
\newcommand{\cmark}{\ding{51}}
\newcommand{\xmark}{\ding{55}}
\newcommand{\mmark}{\(\pm\)}

\newenvironment{compacttbl}{%
  \begingroup\scriptsize\setlength{\tabcolsep}{2pt}\renewcommand{\arraystretch}{0.92}%
}{\endgroup}

\usepackage{enumitem}
\usepackage{tablefootnote}

\begin{document}
\title{Comparing Stochastic and Ray-tracing Datasets in Machine Learning for Wireless Applications}

\author{João Morais\textsuperscript{1}, Akshay Malhotra\textsuperscript{2}, Shahab Hamidi-Rad\textsuperscript{2}, Ahmed Alkhateeb\textsuperscript{1} \\
\textsuperscript{1} Wireless Intelligence Lab, Arizona State University, USA \ \{joao, alkhateeb\}@asu.edu \\
\textsuperscript{2} InterDigital, Inc., USA \ \{akshay.malhotra, shahab.hamidi-rad\}@interdigital.com}
\maketitle

\begin{abstract}


Machine learning for wireless systems is commonly studied using standardized \emph{stochastic} channel models (e.g., TDL/CDL/UMa) because of their legacy in wireless communication standardization and their ability to generate data at scale. However, some of their structural assumptions may diverge from real-world propagation. This paper asks when these models are sufficient and when \emph{ray‑traced} (RT) data—a proxy for the real-world—provides tangible benefits. To answer these questions, we conduct an empirical study on two representative tasks: CSI compression and temporal channel prediction. Models are trained and evaluated using in-domain, cross-domain, and small-data fine-tuning protocols. Across settings, we observe that stochastic-only evaluation may over- or under-estimate performance relative to RT. These findings support a task-aware recipe where stochastic models can be leveraged for scalable pre-training and for tasks that do not rely on strong spatiotemporal coupling. When that coupling matters, pre-training and evaluation should be grounded in spatially consistent or geometrically similar RT scenarios. This study provides initial guidance to inform future discussions on benchmarking and standardization.

\end{abstract}

\section{Introduction}\label{sec:intro}

Researchers in academia and industry still train and evaluate many wireless machine-learning (ML) systems on standardized \emph{stochastic} channel models—most prominently the 3GPP TR~38.901 TDL/CDL profiles and scenario families (e.g., UMa, UMi), and widely used geometry-based stochastic models such as WINNER~II, NYUSIM, and QuaDRiGa~\cite{3gpp38901,winner2,NYUSIM, jaeckel2014quadriga}. These models are attractive because they are well documented, reproducible, and implemented in mature tools~\cite{jaeckel2014quadriga,winner2}. Yet their convenience comes with assumptions about spatial/temporal evolution (e.g., how clusters appear/disappear, how Doppler is modeled, how spatial consistency is enforced) that differ from site-specific propagation captured by map-based or ray-tracing data~\cite{wang2018survey,lim2018mapbased}. Those assumptions can be unrealistic for tasks that need physically faithful structure: they can freeze or oversimplify the coupled evolution across delay, angle, and Doppler, which ray-tracing or real-world channels exhibit~\cite{lim2018mapbased,jaeckel2018sos,kurras2018sc}. 

Many ML papers, therefore, produce plausible results on synthetic channels while the training data quietly shapes what the models can and cannot learn. The risk often goes unnoticed because stochastic generators yield clean, abundant, and internally consistent samples; the discrepancy becomes visible only when we compare to measured or site-specific (ray-traced) channels~\cite{lim2018mapbased}. At the same time, the practical barrier to deterministic data has dropped: public ray-tracing (RT) datasets and differentiable RT simulators (e.g., DeepMIMO; SionnaRT) now make site-specific channels nearly as easy to obtain and manipulate as 38.901-compliant outputs~\cite{alkhateeb2019deepmimo,hoydis2023sionnaRT}. In short, data that is easy to generate can quietly undermine research conclusions when its modeling choices do not match the task’s information needs.

\paragraph*{Scope} This paper is an \emph{initial}, task-focused study—not a comprehensive guideline. We assemble basic criteria for when common stochastic models (TDL, CDL, UMa) are appropriate for ML tasks and when they are not, and we juxtapose them with ray-traced alternatives. We offer this as a starting point for public guidance that future research can refine. 

We study two representative ML problems that stress different channel characteristics. The first is \emph{CSI compression}, where autoencoders learn a low-rate representation for feedback; all channel model families generate compatible multi-dimensional channels once normalized, so stochastic and ray-traced data are both viable. The second is \emph{channel prediction} from a fixed input window using a GRU-based model. Here, stochastic profiles typically constrain temporal evolution via Doppler statistics and stationary clusters, whereas site-specific channels exhibit coupled delay–angle–Doppler dynamics that a predictor can exploit. Consequently, models trained only on stochastic data face a larger domain shift and generalize less reliably to site-specific channels. To assess this, we (i) train and test using data from each channel model, (ii) cross-test between stochastic and ray-traced models, and (iii) pre-train and fine-tune across channel models to measure transfer.

\begin{figure*}[t]
\centering
\resizebox{\textwidth}{!}{%
\begin{tikzpicture}[>=Stealth, font=\small]

\fill[gray!10] (0,1.8) rectangle (12.0,3.4);    
\fill[gray!10] (12.0,1.8) rectangle (25.0,3.4); 
\fill[gray!10] (25.0,1.8) rectangle (31.2,3.4); 

\node[align=center, font=\bfseries\Large] at (6.0,2.6)  {Purely Stochastic};
\node[align=center, font=\bfseries\Large] at (17.0,2.6) {Spatial Channel Models};
\node[align=center, font=\bfseries\Large] at (27.1,2.6) {Map-Based};

\fill[gray!65] (0,0) rectangle (30.8,0.7);
\fill[gray!65] (30.8,-0.25) -- (31.2,0.35) -- (30.8,0.95) -- cycle;
\node[white, font=\bfseries\Large] at (16,0.35) {Determinism};

\newcommand{\ticktop}[2]{%
  \draw[thick] (#1,0.70) -- (#1,1.10);
  \node[align=center, anchor=base, font=\large] at (#1,1.25) {#2};
}

\ticktop{1.2}{AWGN}
\ticktop{3.2}{Rayleigh}
\ticktop{4.9}{Rician}
\ticktop{7.0}{Nakagami-$m$}
\ticktop{8.9}{TDL}  

\ticktop{10.6}{CDL}

\ticktop{12.9}{UMi/UMa/RMa}

\ticktop{15.7}{WINNER~II}
\ticktop{18.2}{COST~2100}
\ticktop{20.3}{NYUSIM}
\ticktop{22.4}{QuaDRiGa}

\ticktop{25.7}{CDL w/ RT clusters}
\ticktop{29.5}{Ray Tracing}

\draw[dashed, thick] (9.8,-0.4) -- (9.8, 1.0);
\node[anchor=base east, font=\normalsize] at (9.8, -0.4) {no array structure};
\node[anchor=base west, font=\normalsize] at (9.8, -0.4) {array structure \& polarization coupling};

\draw[dashed, thick] (23.6,-0.4) -- (23.6, 1.0);
\node[anchor=base east, font=\normalsize] at (23.6, -.4) {random clusters};
\node[anchor=base west, font=\normalsize] at (23.6, -.4) {fixed clusters computed from scene};

\node[anchor=base, align=left, scale=1.5] at (1.2, -0.62) {$y = x + n$};
\node[anchor=base, align=left, scale=1.5] at (3.5, -0.62) {$y = h x + n$};
\node[anchor=base, align=left, scale=1.5] at (2.2,-1.62) {$n,h \sim \mathcal{CN}(0,\sigma^2)$};
\node[anchor=base, align=left, scale=1.5] at (6.2,-1.3) {$R \sim \mathrm{Rician}(K,\Omega)$};
\node[anchor=base, align=left, scale=1.5] at (6.55,-1.9) {$R \sim \mathrm{Nakagami}(m,\Omega)$};

\node[anchor=north west, align=left, scale=1.5] at (8.99,-0.85)
{$\displaystyle H(t,f)=\sum_{\ell=1}^{L}\alpha_{\ell}\,
   e^{-j2\pi f \tau_{\ell}}\, e^{j2\pi f_{D,\ell} t}$};

\draw[decorate, decoration={brace, mirror, amplitude=6pt}] (10.2,-0.62) -- (30.0,-0.62);
\node[anchor=north, align=center, scale=1.5] at (22.5,-0.78)
{$\displaystyle \mathbf{H}(t,f)=\sum_{n=1}^{N_{\text{paths}}} \alpha_n\,
\mathbf{a}_{\mathrm{rx}}(\theta^{\mathrm{rx}}_n)\,
   \mathbf{a}_{\mathrm{tx}}^{\!H}(\theta^{\mathrm{tx}}_n)\,
   e^{j(2\pi f \tau_n - 2\pi f_{D,n} t)}$};

\end{tikzpicture}}
\vspace{-1.3\baselineskip} 
\caption{Determinism continuum with two explicit boundaries. The first boundary marks where \emph{array manifold} modeling becomes usable (CDL and beyond). The second boundary separates \emph{random/flexible clusters} from \emph{fixed clusters} derived from ray-tracing. Spatial Channel Models and Map‑Based (Ray Tracing) share the same geometric form; they differ in how path parameters are generated. 
Parameters: noise variance $\sigma_n^2$, Rician $K$-factor; average power $\Omega$; Nakagami $m$; complex gains $\alpha_{\ell/n}$, delays $\tau_{\ell/n}$, and doppler shifts $f_{D,\ell/n}$, per tap $\ell$ or path $n$; path angles of arrival/departure $\theta^{rx/tx}_n$; array response vectors $\mathbf{a}_{\mathrm{rx/tx}}$}
\label{fig:determinism_continuum_full}
\end{figure*}
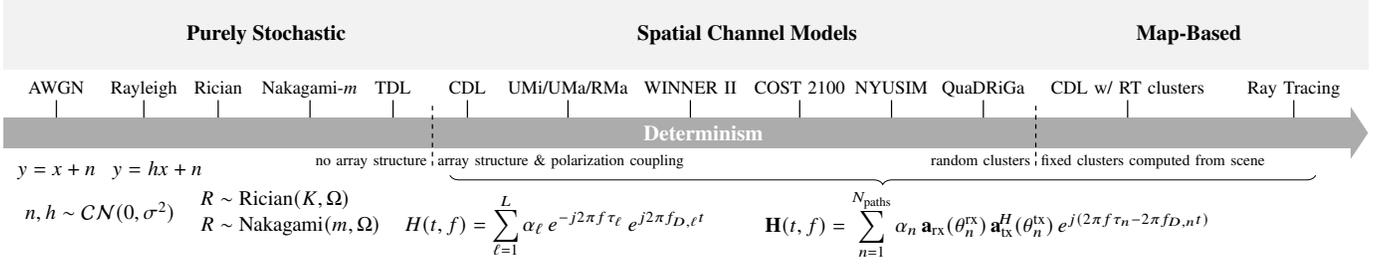

\textbf{We attempt to answer three concrete questions:}
\begin{itemize}
  \item \textbf{Q1: Sufficiency.} If we train and evaluate only on stochastic data, do we obtain performance expectations that fail to hold on site-specific (ray-traced) data?
  \item \textbf{Q2: Relative utility.} When (by task and conditions) does ray-traced data deliver meaningfully better training and evaluation signals than stochastic data?
  \item \textbf{Q3: Model selection.} For a given ML task, which stochastic model class (TDL, CDL, UMa) is theoretically suitable given its assumptions, and when should we avoid it in favor of site-specific data?
\end{itemize}


\section{Background: Channel Models and Consistency} \label{sec:background}

\subsection{A concise taxonomy}
We group channel generators by how they encode geometry and evolution.
\textbf{Purely stochastic} models (e.g., TDL/CDL and standardized scenarios UMa/UMi in 3GPP TR~38.901) specify taps, powers, and Doppler statistics with limited or no explicit scene geometry~\cite{3gpp38901}.
\textbf{Geometry-based stochastic models (GBSMs)} (WINNER~II, COST~2100, NYUSIM, QuaDRiGa) attach stochastic clusters to simple geometric constructs to induce spatial correlation and time evolution~\cite{winner2,liu2012cost2100,jaeckel2014quadriga,NYUSIM,7996792}.
\textbf{Deterministic} (map-/ray-tracing) models derive multipath from site geometry and EM interactions, enabling scene-specific behavior~\cite{lim2018mapbased,zemen2024sscr}.
This work relies on these families to avoid rederiving well-known equations and focus instead on the less explicit modeling choices that matter for machine learning.

\subsection{Implicit modeling choices that affect ML outcomes}
\paragraph*{(C1) Time--frequency consistency}
TDL/CDL impose frequency correlation through tap delays and time evolution mainly through per-path Doppler processes; phase evolves coherently within each path but without explicit delay--angle coupling beyond what the profile prescribes~\cite{3gpp38901}. RT/map-based models preserve joint delay--angle--Doppler structure induced by geometry and motion~\cite{lim2018mapbased,zemen2024sscr}. \emph{Implication:} tasks that learn cross-sample dynamics (e.g., prediction, tracking) may benefit from models that preserve coupled evolution.

\paragraph*{(C2) Spatial consistency and multi-user link correlation}
Spatial consistency in 38.901 is optional and implemented via correlated random fields or sum-of-sinusoids methods~\cite{jaeckel2018sos,kurras2018sc}. GBSMs like COST~2100 also use common clusters and visibility regions to correlate links~\cite{liu2012cost2100}. \emph{Implication:} beam tracking, handover, and multi-user tasks would need spatially correlated channels; otherwise, models may oversimplify dependence.

\paragraph*{(C3) Cluster birth/death and LOS/NLOS transitions}
GBSMs simulate cluster appearance, disappearance, and scenario transitions through stochastic rules; realism depends on parameterization~\cite{winner2,liu2012cost2100}. RT transitions arise from geometry and occlusion. \emph{Implication:} abrupt changes (corners, blockages) are hard to capture with smooth stochastic evolutions.

\paragraph*{(C4) Array manifold and field region}
Most standardized stochastic profiles assume far-field array responses; per-antenna path-length differences and mutual coupling are typically simplified. RT engines can model spherical wavefronts and per-element variations when needed~\cite{lim2018mapbased}. \emph{Implication:} near-field beam focusing, very large arrays, and polarization-sensitive tasks need care when using stochastic models.

\paragraph*{(C5) Frequency dependence and materials}
Stochastic models rarely account for frequency-dispersive materials. Modern RT frameworks can incorporate material properties and even differentiate with respect to them~\cite{hoydis2023sionnaRT}. \emph{Implication:} cross-band generalization, RIS/material inference, and joint comm–sensing tasks often require site-specific models.

\paragraph*{(C6) Data availability and reproducibility}
Historically, stochastic generators dominated because they were easy to run and reproduce. Today, public RT datasets and tools (e.g., DeepMIMO; Sionna~RT) make site-specific channels accessible with scriptable APIs~\cite{alkhateeb2019deepmimo,hoydis2023sionnaRT}. \emph{Implication:} the practicality gap has narrowed; selection should be driven by task assumptions rather than tool friction.

\subsection{From channel models to ML task requirements}

Fig.~\ref{fig:determinism_continuum_full} shows how channel families move from amplitude/noise statistics to array- and geometry-aware models, along with the governing formulas. These equations make the channel assumptions of each model more explicit. Model assumptions/approximations come from how the model derives the quantities in the geometric equation, and how closely they depend on the environment. To assess whether ray-tracing data is necessary, we must determine whether such approximations are good or bad, which depends on the application.  Therefore, we elected two tasks in this study: One where most approximations seem harmless, and one where they may result in differences in the application outcome.

\paragraph*{Focus tasks in this study} We study channel/CSI compression and prediction. \textbf{CSI compression} is broadly feasible across model families once channels are normalized, thereby probing how realism affects compression and cross-domain transfer when the basic ingredients are present in all datasets. \textbf{Temporal channel prediction} stresses coupled delay–angle–Doppler evolution and spatial consistency; several stochastic profiles lack these couplings by construction, so this task exposes domain shift most clearly. Section~\ref{sec:data} describes the generation of stochastic and ray-tracing data, and Section~\ref{sec:methods} defines architectures and protocols for in-domain training, cross-testing, and pre-training \& fine-tuning.

\section{Data Generation: Stochastic \& Ray-tracing}\label{sec:data}
This section describes the data generation process—covering what data is produced, why it is generated in this manner, and how these choices support the two tasks evaluated later. To ensure channel model comparability, we construct channels equally per model and learning pipeline, so that any performance differences arise from model assumptions rather than data handling.

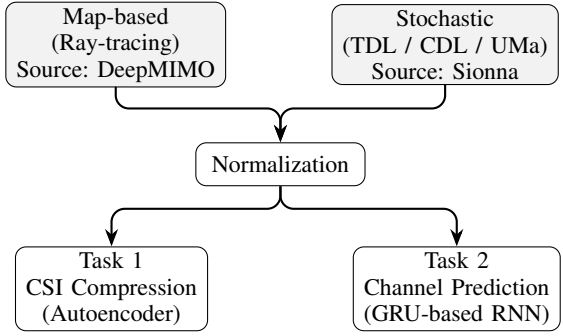
\begin{figure}[t]
\centering
\begin{tikzpicture}[
  >=Stealth,
  every node/.style={font=\small},
  box/.style={
    draw, rounded corners, fill=gray!10,
    align=center, inner xsep=4pt, inner ysep=2pt
  },
  sbox1/.style={
    draw, rounded corners, 
    align=center,              
    inner xsep=6pt, inner ysep=5pt 
  },
  sbox2/.style={
    draw, rounded corners, 
    align=center,              
    inner xsep=4pt, inner ysep=2pt 
  }
]

\node[box] (map) {Map-based\\(Ray-tracing)\\Source: DeepMIMO};
\node[box, right=14mm of map] (stoch) {Stochastic\\(TDL / CDL / UMa)\\Source: Sionna};

\node[sbox1, below=13mm of $(map)!0.5!(stoch)$] (norm) {Normalization};

\node[sbox2, below left=8mm and -1.7mm of norm] (t1) {Task 1\\CSI Compression\\(Autoencoder)};
\node[sbox2, below right=8mm and -1.7mm of norm] (t2) {Task 2\\Channel Prediction\\(GRU-based RNN)};

\coordinate (tap) at ($(norm.north)+(0,4.5mm)$);
\draw[->, thick, rounded corners=5pt, line cap=round]
  (map.south)  |- (tap) -- (norm.north);
\draw[->, thick, rounded corners=5pt, line cap=round]
  (stoch.south)|- (tap) -- (norm.north);

\coordinate (split) at ($(norm.south)+(0,-3mm)$);
\draw[->, thick, rounded corners=5pt]
  (norm.south) -- ++(0,-3mm) -| (t1.north);
\draw[->, thick, rounded corners=5pt]
  (norm.south) -- ++(0,-3mm) -| (t2.north);

\end{tikzpicture}
\caption{Data-to-model flow. Two generators feed task-specific normalization, following the ML models: autoencoder (Task~1) or GRU (Task~2). Differences in data generation between Task~1 and 2 are: number of antennas (32 vs 2), number of subcarriers (32 vs 1), sequence length (1 vs $L$), and Doppler progression (only in Task~2, applied along sequence). }
\label{fig:data_pipeline}
\end{figure}

Datasets come from two sources. Stochastic datasets use Sionna PHY~\cite{sionna} with 3GPP‑compliant TDL, CDL, and UMa profiles. Map‑based datasets use DeepMIMO~\cite{alkhateeb2019deepmimo}\footnote{DeepMIMO scenes were ray‑traced with Wireless InSite~\cite{wireless_insite} and are freely available (150+ scenarios) at \href{https://deepmimo.net}{https://deepmimo.net}.}. Figure~\ref{fig:data_pipeline} shows both branches feeding a shared preprocessing path and then the two models; we keep tensor conventions identical within each task to isolate the effect of the channel generator.

\vspace{-.2cm}

\subsection{Data sources: Sionna \& DeepMIMO}
\textbf{Stochastic (Sionna—TDL/CDL/UMa).} For Task~1 (compression), we synthesize wideband, multi‑antenna snapshots without Doppler. For Task~2 (prediction), we generate sequences with Doppler and enable spatial consistency where available; fixed‑length windows serve as inputs and targets. The data for both tasks comes from the stochastic channel model implementations in \texttt{sionna.phy.channel.tr38901}.

\textbf{Map‑based (DeepMIMO—Ray-tracing).} We use site‑specific scenes. For compression, we sample snapshots at fixed spatial spacing; for prediction, we follow straight‑line user tracks and interpolate path components (powers, phases, angles, delays) to match the desired inter‑sample distances. Table \ref{tab:rt_params} contains the specific ray-tracing parameters for each task. The RT scenarios differ per task; Task~1 compares across three \(1\,\mathrm{m}\)-resolution scenarios with multiple BSs, on which we aggregate all BS-UE links totalling over 50,000 receivers per scenario; Task~2 uses an ultra-dense \(10\,\mathrm{cm}\)-resolution version of the ASU campus scenario with 12 million receivers, which is necessary to obtain channels close enough for channel prediction. We simulate all propagation phenomena, except penetrations. The high-density scenario in Task~2 does not include scattering events, only diffraction and reflections.

\subsection{Preprocessing and normalization}  
\textbf{Common settings}: 
For all experiments, we follow the channel modeling parameters defined in the 3GPP technical specification 38.901 \cite{3gpp38901}. We include key common settings used across models and tasks in Table~\ref{tab:common_params}. Among these, we use a stochastic delay spread of \(300\)~ns to provide a richer multipath profile. The simulated arrays adopt a uniform linear array (ULA) configuration with half-wavelength (\(\lambda/2\)) antenna spacing, a well-established choice. Spatial correlation is enabled according to the specification, and the simulations operate in the classical FR1, sub-6~GHz band with a suitable 5G numerology. No additive noise is applied, effectively setting the signal-to-noise ratio (SNR) to infinity to focus on the intrinsic properties of the channel.  

\textbf{Normalization and subcarrier generation.} Task‑specific normalization differs by design. For \emph{compression} (Task~1), we apply zero‑mean/unit‑variance (ZMUV) normalization per snapshot. To obtain \(32\) effective subcarriers, we generate multiple tones per PRB and \emph{average within each PRB}, selecting \(32\) PRBs to yield \(32\) averaged subcarriers. We then transform to angle–delay by FFT across frequency (delay) and across the array (angle), and \emph{trim the first 16 delay taps}. For \emph{prediction} (Task~2), we form sequences and apply max‑absolute scaling (divide by the dataset’s \(\max |H|\)); a single subcarrier is used to focus learning on temporal evolution.

\begin{table}[t]
\centering
\caption{Common parameters used across datasets and tasks.}
\label{tab:common_params}
\small
\begin{tabular}{l l}
\toprule
Subcarrier spacing ($\Delta f$) & $15\,\mathrm{kHz}$ \\
Carrier frequency ($f_c$) & $3.5\,\mathrm{GHz}$ \\
Stochastic delay spread & $300\,\mathrm{ns}$ \\
Arrays & ULA, $\lambda / 2$ spacing \\
Bandwidth & $N_{\text{sc}}\times \Delta f$ \\
DeepMIMO version & 4.0.0b11 \\
Sionna version & 0.19.2 \\
Reproducibility & Code Open-source \tablefootnote{Available in \href{https://github.com/jmoraispk/StochasticRTcomparison}{https://github.com/jmoraispk/StochasticRTcomparison}} \\
\bottomrule
\end{tabular}
\end{table}

\begin{table}[t]
\centering
\caption{Map-based (ray-tracing) parameters per task.}
\label{tab:rt_params}
\small
\begin{tabular}{p{1.3cm} p{3.15cm} p{3.05cm}}
\toprule
 & \textbf{Task 1: Compression} & \textbf{Task 2: Prediction} \\
\midrule
Scenes & \texttt{\footnotesize asu\_campus\_3p5} \texttt{\footnotesize city\_0\_newyork\_3p5} \texttt{\footnotesize city\_1\_losangeles\_3p5} & \texttt{\footnotesize asu\_campus\_3p5\_10cm} \\
Spacing & $1\,\mathrm{m}$ & $10\,\mathrm{mm}$ (interpolated) \\
Reflections & $5$ & $5$ \\
Diffractions & $1$ & $1$ \\
Scattering & $1$ & $0$ \\
Penetration & No & No \\
\bottomrule
\end{tabular}
\end{table}

\subsection{Processing Differences Between Tasks}
We highlight how the two tasks diverge operationally. Table~\ref{tab:task_specific_params} lists the settings that are changed between tasks, and we explain \emph{why} they differ and how those choices shape the tensors consumed by the models in Section~\ref{sec:methods}.

\paragraph{What changes and why.}
(i) \textbf{Sample count:} Task~1 (compression) uses $50{,}000$ snapshots, while Task~2 (prediction) uses $200{,}000$ sequences. The predictor benefits more from data because it must learn temporal dynamics, while the autoencoder reaches stable performance with fewer samples. (ii) \textbf{Subcarriers and antennas:} Task~1 uses $32$ subcarriers and a larger array ($N_t{=}32$) to exploit within‑snapshot structure. Task~2 uses a single subcarrier and a small array ($N_t{=}2$) to isolate temporal evolution, and limit compute complexity and dataset size, since matrix dimensions will have a new time axis with up to $L{=}60$ samples. (iii) \textbf{Sampling, Doppler, and speed:} Sampling period, Doppler settings, and user speeds apply \emph{only} to Task~2, because only prediction spans time. This additional processing needed for Task~2 is detailed ahead.

\begin{table}[t]
\centering
\caption{\hspace{-.05cm}Task-specific parameters for channel compression \& prediction}
\label{tab:task_specific_params}
\small
\setlength{\tabcolsep}{5pt} 
\begin{tabular}{l p{2.9cm} p{2.5cm}}
\toprule
 & \textbf{Task 1: Compression} & \textbf{Task 2: Prediction} \\
\midrule
Samples ($N$) & $50{,}000$ & $200{,}000$ \\
Subcarriers ($N_c$) & $32$ & $1$ \\
Tx Antennas ($N_t$) & $32$ & $2$ \\
Seq. length ($L$) & 1 (single slot) & $60$ \\
Sampling period & N/A & $1\,\mathrm{ms}$ \\
Doppler ($f_D$) & N/A & $100\,\mathrm{Hz}$ \\
Speed & N/A & $\approx30\,\mathrm{km/h}$ \\
\bottomrule
\end{tabular}
\vspace{-.5cm}
\end{table}

\paragraph{Doppler modeling (Task~2 only).} In 
\textbf{stochastic sequences}, 3GPP‑compliant generators assign each subpath an angle drawn from cluster distributions and compute per‑path Doppler by projecting the (randomized) motion direction onto that random angle; phase then evolves consistently across the window while the large-scale parameters of each cluster remain constant. The result is a constant Doppler phase progression, without angle-delay channel progressions. In \textbf{Map‑based sequences}, angle-delay channel evolution exists by design. Realistic modeling of Doppler effects can be achieved using sequence-specific movement speeds and orientations. That orientation and speed magnitude, when projected onto the user paths via the angles of arrival/departure at the user, produces a Doppler shift per path. The shifts per path are then superimposed in the geometric channel equation. The difference to the stochastic approach is that the angles at the receiver and the mobility profiles are not random, but determined by the scene geometry. Such Doppler modeling is supported in DeepMIMO. Finally, the ray-tracing inter-sample spacing and Doppler are chosen to be mutually consistent with the simulated speeds at $3.5\,\mathrm{GHz}$.  

\paragraph{Sequence construction \& interpolation (Task~2 only).}
User trajectories follow straight lines (constant $x$ or constant $y$). The base ray-tracing simulation uses a $10\,\mathrm{cm}$ grid. Therefore, to achieve $10\,\mathrm{mm}$ spacings, we \emph{linearly interpolate} per‑path components (receiver positions, angles, powers, phases, and delays) along the track. Linear interpolation is typically sufficient when inter-sample distances are smaller than the wavelength $\lambda_{3.5\,\mathrm{GHz}} \approx 86 mm$. These interpolated sequences span $L{=}60$ samples to support any combination of input sequence sizes and output prediction horizons that span 60 ms, i.e., the first sample of the input sequence and the last sample of the output are within 60 ms.

\section{Machine Learning Models for Channel Compression \& Channel Prediction}\label{sec:methods}

To understand the impact of stochastic versus ray-tracing channel modeling, we evaluate two tasks: \textbf{CSI compression} with a CSINet+ variant, and \textbf{temporal channel prediction} with a GRU-based predictor. This section specifies the machine learning models used in the tasks, the concrete task objectives, and the evaluation protocols used to answer the questions posed in Section~\ref{sec:intro}. The previous section focused on settings for data generation and processing; here, we focus on how the data is utilized.

\textbf{Performance metric}: Both tasks optimize normalized mean squared error in decibels, according to
\[
\text{NMSE}_{\mathrm{dB}}=10\log_{10}\!\frac{\lVert \mathbf{H}-\hat{\mathbf{H}}\rVert_2^2}{\lVert \mathbf{H}\rVert_2^2}.
\]
The choice is made because both tasks involve direct channel comparisons: in Task~1, the goal is to compress the channel and compare it to its reconstructed version, whereas in Task~2 the predicted channels are compared against ground truth. Additionally, the channels are split into real-valued components \([\Re,\Im]\) before being used for learning.

\begin{table}[t]
\centering
\begin{compacttbl}
\caption{Task~1 (Channel Compression) Model based on CsiNet+ \cite{wen2017csinet}.}
\label{tab:csinet_layers}
\begin{tabular}{l l l}
\toprule
Stage & Layer & Output shape \\
\midrule
Encoder & Conv2D(2$\to$2, $7\!\times\!7$) + BN + LeakyReLU & $(B,2,N_c,N_t)$ \\
       & Conv2D(2$\to$2, $7\!\times\!7$) + BN + LeakyReLU & $(B,2,N_c,N_t)$ \\
       & Flatten $\to$ FC $\to$ code (dim $=32$) & $(B,32)$ \\
\midrule
Decoder & FC $\to$ reshape to $(B,2,N_c,N_t)$ & $(B,2,N_c,N_t)$ \\
       & Conv2D(2$\to$2, $7\!\times\!7$) + BN + Tanh & $(B,2,N_c,N_t)$ \\
       & RefineNet $\times 6$ (Conv $7\!\times\!7$, $5\!\times\!5$, $3\!\times\!3$ with skip) & $(B,2,N_c,N_t)$ \\
       & L2 norm of output (stability) & $(B,2,N_c,N_t)$ \\
\bottomrule
\end{tabular}
\end{compacttbl}
\end{table}

\subsection{Task 1 — CSI compression with CsiNet+}

\paragraph{Input and trimming.}
Angle–delay tensors from Section~\ref{sec:data} are fed to the network after a frequency$\to$delay FFT and array$\to$angle FFT; Here, only the first 16 delay taps are retained, since they contain most of the channel power. With $N_t{=}32$ TX antennas and two real channels (I/Q), the input has shape $(B,2,N_c{=}16,N_t{=}32)$, i.e. $32\!\times\!16\!\times\!2=1024$ features. The encoded dimension is fixed at 32 (compression rate 32).

\paragraph{Architecture.}
The model uses two $7{\times}7$ convolutional blocks in the encoder and a mirrored decoder followed by \textbf{six} refinement nets (ResNet-style). The output of the decoder is $\ell_2$-normalized for stability. BatchNorm (BN) is also used to prevent overfitting. Table \ref{tab:csinet_layers} details the architecture, which is also open-source with the code.

\paragraph{Training and protocols.}
We train the model for \textbf{50} epochs with batch \textbf{128} and learning rate \textbf{$10^{-2}$}. Then, we repeat each experiment \textbf{10} times and report the mean. Each run uses an \textbf{80/10/10} split (train/val/test) within the source domain. Evaluate three protocols:  
(i) \emph{In-domain}: train and test within the same channel model.  
(ii) \emph{Cross-testing}: take the in-domain trained model and evaluate it on \textbf{100\%} of each target dataset.  
(iii) \emph{Fine-tuning}: update \emph{all} layers of the model on a percentage of the target domain and report performance on the remainder.  
Additionally, to assess the impact of pre-training, we use fractions of the total available data (e.g. \textbf{[0.5, 1, 5, 10, 40, 90]\%} of the $85$k samples of the ASU scenario) to compare the performance of in-domain trained models with pre-trained models fine-tuned. This complementary experiment aims to understand the utility of stochastic models for pre-training.

\subsection{Task 2 — Temporal channel prediction with a GRU}

\begin{table}[t]
\centering
\begin{compacttbl}
\caption{Task~2 (Channel Prediction) Model based on GRUs \cite{gru_pred_mdpi}.}
\label{tab:gru_layers}
\begin{tabular}{l l l}
\toprule
Stage & Layer & Output shape \\
\midrule
Encoder & GRU($\text{in}=4$, hidden$=64$, layers$=2$, dropout$=0.5$) & $(B,T,64)$ \\
Head    & Take last step $\to$ BN1d(64) $\to$ FC(64$\to$4) & $(B,4)$ \\
\bottomrule
\end{tabular}
\end{compacttbl}
\end{table}

\paragraph{Input windows and horizons.}
Sequences have total length \textbf{$L{=}60$}: the first \textbf{20} samples are inputs ($L_{\mathrm{in}}{=}20$); the next \textbf{40} are targets at \textbf{\{1, 3, 5, 10, 20, 40\}}\,ms ahead of the last input sample. Note, the model output is single-sample ($L_{\mathrm{out}}{=}1$). 

\paragraph{Architecture.}
A two-layer GRU (hidden 64, dropout 0.5) processes each window; the last hidden state passes through BatchNorm1d and a linear head. The per-step feature size equals the flattened real–imag channel (2) times antennas (2) and subcarriers (1), which equals 4. Table~\ref{tab:gru_layers} details further the model architecture.

\paragraph{Training and protocols.}
The model trains for \textbf{300} epochs with batch \textbf{256} and initial learning rate \textbf{$4{\times}10^{-4}$}. Protocols mirror Task~1 (in-domain, cross-testing, fine-tuning) and use the same \textbf{80/10/10} split. Table \ref{tab:learn_hparams} summarizes these hyperparameters. Additionally, validation loss is monitored to trigger an early stop if it has not improved for 60 epochs.

\begin{table}[h]
\centering
\caption{Learning hyperparameters for each task/model.}
\label{tab:learn_hparams}
\begin{tabular}{l c c c c c c}
\toprule
Task & LR & Batch & Epochs & Repeats & Split \\
\midrule
T1 (CsiNet+) & $1\!\times\!10^{-2}$ & 128 & 50 & 10 & 80/10/10 \\
T2 (GRU)     & $4\!\times\!10^{-4}$ & 256 & 300 & 10 & 80/10/10 \\
\bottomrule
\end{tabular}
\end{table}

\begin{table*}[ht]
\centering
\caption{Cross-Test and Fine-Tuning (1\%) Results (NMSE in dB)}
\label{tab:cross_finetune}
\renewcommand{\arraystretch}{1.1}
\begin{tabular}{lccccccc|ccccccc}
\toprule
& \multicolumn{7}{c|}{\textbf{Cross-Test}} & \multicolumn{7}{c}{\textbf{Fine-Tuning (1\%)}} \\
\cmidrule(lr){2-8} \cmidrule(lr){9-15}
\textbf{Training} &
\textbf{TDL-C} &
\textbf{CDL-C} &
\textbf{CDL-D} &
\textbf{UMa} &
\textbf{NY} &
\textbf{LA} &
\textbf{ASU} &
\textbf{TDL-C} &
\textbf{CDL-C} &
\textbf{CDL-D} &
\textbf{UMa} &
\textbf{NY} &
\textbf{LA} &
\textbf{ASU} \\
\midrule
TDL-C   & $\mathbf{-6.3}$  & $-11.0$& $-4.8$ & $-4.9$ & $-10.6$ & $-10.9$ & $-5.5$
        & $\mathbf{-6.3}$  & $-18.6$& $-12.6$& $-4.9$ & $-16.5$ & $-14.5$ & $-13.1$ \\
CDL-C   & $-1.8$  & $\mathbf{-27.3}$& $-2.9$ & $-1.2$ & $-4.5$  & $-4.5$  & $-3.2$
        & $-5.2$  & $\mathbf{-27.3}$& $-14.5$& $-3.7$ & $-11.6$ & $-12.9$ & $-11.1$ \\
CDL-D   & $-0.1$  & $-0.1$ & $\mathbf{-18.8}$& $0.5$  & $-1.3$  & $-2.9$  & $0.4$
        & $-5.3$  & $-18.8$& $\mathbf{-18.8}$& $-3.7$ & $-12.3$ & $-11.4$ & $-9.7$ \\
UMa     & $-3.9$  & $-10.7$& $-6.4$ & $\mathbf{-6.2}$ & $-13.1$ & $-12.6$ & $-7.7$
        & $-5.2$  & $-18.7$& $-13.4$& $\mathbf{-6.2}$ & $-16.1$ & $-15.2$ & $-14.6$ \\
NY (RT) & $-4.4$  & $-12.8$& $-5.3$ & $-4.2$ & $\mathbf{-22.7}$ & $-21.4$ & $-8.5$
        & $-5.1$  & $-18.1$& $-14.2$& $-4.6$ & $\mathbf{-22.7}$ & $-20.9$ & $-16.7$ \\
LA (RT) & $-4.5$  & $-11.7$& $-5.1$ & $-4.1$ & $-18.9$ & $\mathbf{-23.5}$ & $-7.0$
        & $-5.0$  & $-16.9$& $-14.5$& $-4.3$ & $-19.9$ & $\mathbf{-23.5}$ & $-15.4$ \\
ASU (RT)& $-2.1$  & $-8.7$ & $-4.3$ & $-3.8$ & $-14.1$ & $-13.0$ & $\mathbf{-21.3}$
        & $-4.2$  & $-17.9$& $-13.3$& $-4.4$ & $-16.2$ & $-16.3$ & $\mathbf{-21.3}$ \\
\bottomrule
\end{tabular}
\end{table*}

\section{Results \& Discussion}


\subsection{Task~1: Channel Compression}

\paragraph{\textbf{In-Domain Evaluation}}
Table \ref{tab:cross_finetune} exhibits the expected diagonal dominance---training and testing within the same domain typically minimizes NMSE---yet with systematic exceptions that are informative. In particular, TDL-C and UMa show poor in-domain learnability ($\approx -6$ dB), while paradoxically transferring better to other domains (e.g., TDL-C$\to$CDL-C $-11.0$ dB; UMa$\to$NY $-13.1$ dB). This pattern suggests that these two stochastic models induce broader, higher-variance channel distributions with limited spatial consistency (e.g., wider spreads over delay/angle and weaker spatial correlations) that are difficult to compress faithfully, yet their learned features remain useful as generic priors elsewhere. Among RT scenarios, LA and NY set the strongest in-domain baselines ($-23.5$ dB and $-22.7$ dB), with ASU slightly weaker ($-21.3$ dB) but, as shown below, ASU's representations transfer outward well. Among stochastic models, CDL-C is a conspicuous ``easy'' case ($-27.3$ dB), reflecting its more concentrated multipath structure and higher regularity compared to TDL-C and UMa.

\paragraph{\textbf{Cross-testing (zero-shot)}}
Zero-shot transfer is dominated by geometric similarity. The RT triad exhibits the best off-diagonals overall (NY$\to$LA $-21.4$ dB; LA$\to$NY $-18.9$ dB), while RT$\to$stochastic and stochastic$\to$stochastic transfers are markedly weaker on average (off-diagonal means around $-6$ dB to $-4$ dB). A notable directional asymmetry emerges: ASU$\to$NY/LA achieves $-14.1/-13.0$ dB, whereas NY/LA$\to$ASU drops to $-8.5/-7.0$ dB. This is consistent with ASU being a larger, more scatter-rich campus whose learned embeddings cover more propagation conditions, while models trained on denser urban canyons (NY/LA) under-represent ASU's variability. Within the stochastic family, UMa transfers better to RT than TDL-C ($-11.13$ dB vs.\ $-9.00$ dB on average to ASU/NY/LA), plausibly because UMa retains coarse spatial antenna-structures absent in TDL-C. Moreover, both UMa and TDL-C transfer much better than CDL-C/D, likely due to their greater cluster diversity; CDLs are comparatively constrained and therefore less generalizable.

\paragraph{\textbf{Fine-tuning (1\%)}}
With only 1\% target data, transfer quality improves by $\approx 5.6$ dB on average across off-diagonals, and the largest gains occur where the source representations must be realigned to the target's geometric/statistical structure. RT$\leftrightarrow$RT asymmetries largely collapse---especially toward ASU (NY$\to$ASU $+8.2$ dB; LA$\to$ASU $+8.4$ dB). In addition, models fine-tuned in CDL-C tend to have strong performance (e.g., ASU$\to$CDL-C $+9.2$ dB; TDL-C$\to$CDL-C $+7.6$ dB), further indicating that CDL-C is a specific and comparatively easy scenario. Interestingly, once fine-tuned, the best average starting points (row-wise means over off-diagonals) are UMa ($-13.87$ dB) and TDL-C ($-13.37$ dB), i.e., their broad priors adapt rapidly with small target supervision. However, UMa and TDL-C remain challenging targets (column means $-4.27/-5.00$ dB), indicating that adapting to these stochastic scenarios yields limited results. Practical implication: for deployments in structured environments, pre-training on a scatter-rich RT set (e.g., ASU) or on a broad stochastic prior (UMa/TDL-C) and fine-tuning on $\geq!1$\% of target data reliably closes most of the domain gap; by contrast, purely stochastic evaluation is not a realistic proxy for RT-like channels without adaptation. This low-data fine-tuning assessment is operationally relevant because real-world deployments frequently yield only limited labeled target-domain data.

\begin{figure}[t]
    \centerline{\includegraphics[width=1\columnwidth]{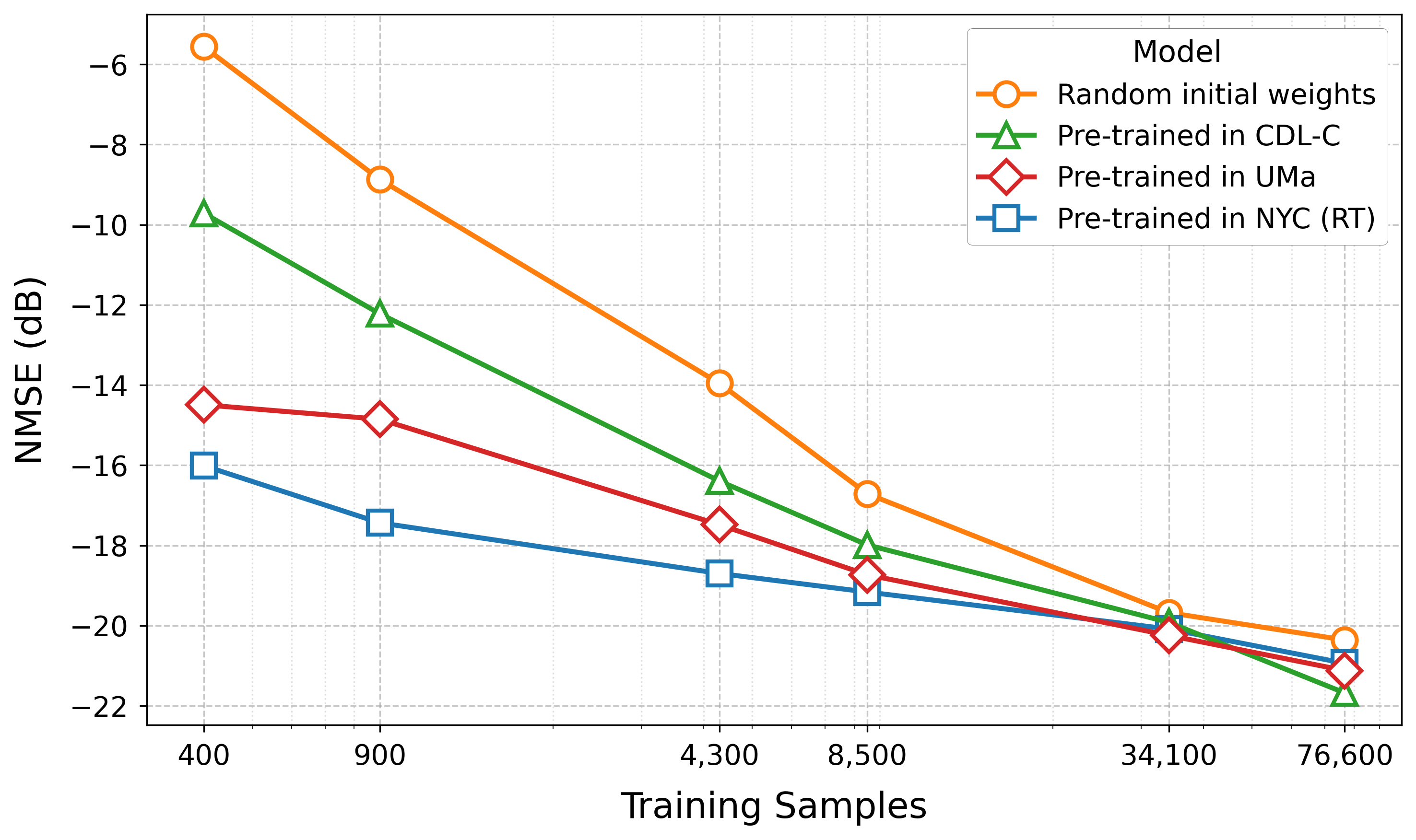}}
    \caption{Channel Compression NMSE on ASU (RT) data as a function of \# training samples (log scale). A randomly initialized model (orange) is contrasted with models pre-trained on 50k samples from CDL-C (green), UMa (red), and NY (RT, blue) and then fine-tuned on the same ray-tracing data.}
    \label{fig:ch_comp}
    \vspace{-.4cm}
\end{figure}

\paragraph{\textbf{The value of pre-training}}
Figure~\ref{fig:ch_comp} shows the NMSE on ASU (RT) as a function of the number of target-domain training samples. It compares a randomly initialized model trained solely on ASU (orange) with models pre-trained on 50k samples from NY (RT, blue), CDL-C (green), and UMa (red), which are then fine-tuned using the same ASU budget. All four curves improve monotonically with more ASU data: the orange baseline moves from $-5.56$~dB at $400$ samples to $-20.36$~dB at $71{,}100$, with pronounced diminishing returns beyond $\approx 31{,}600$ samples (only $0.68$~dB gain thereafter). Pre-training yields substantial gains in the low-data regime and decays toward a small residual advantage as data grows. At $400$ samples, gains over the baseline are $+10.45$~dB (NY), $+8.93$~dB (UMa), and $+4.15$~dB (CDL-C). At $4{,}300$ samples, the advantages remain, ranging from $+4.74$~dB (NY) to $+2.44$~dB (CDL). In summary, when labeled target data are scarce, \emph{pre-training on a geometrically similar RT scene (NY) is most effective} - possibly yielding gains on the order of $\sim\!10$~dB -, with UMa a close second and CDL-C helpful but less aligned. As target data becomes plentiful, all curves converge, and the pre-training effect is typically sub-dB. 

\vspace{-.2cm}

\subsection{Task~2: Channel Prediction}

\paragraph{\textbf{Prediction horizon behavior}}
Figure~\ref{fig:ch_pred} sweeps NMSE (dB) versus horizon for a learned predictor (GRU) and a sample‑and‑hold (SH) baseline. At 1~ms, the GRU curves are tightly clustered across all four datasets (about $-26$ to $-28$~dB) while SH sits near $-7$ to $-10$~dB. By 3–5~ms, the behaviors diverge: the stochastic generators (TDL‑A, CDL‑C, UMa) remain comparatively strong (e.g., at 5~ms roughly $-15.8$, $-17.2$, and $-12.2$~dB on the diagonal of Table~\ref{tab:cross_finetune_stacked}), whereas RT in ASU drops more sharply (around $-8$~dB). Notably, SH crosses into positive NMSE by 5~ms on all datasets. Beyond 10~ms, the stochastic curves approach 0~dB, while ASU exhibits a flatter tail, staying under $-5$~dB until 20~ms. Importantly, similar performance (for example, at 1~ms) does \emph{not} imply similar learned representations; at very short horizons, Doppler dominates, but by a few milliseconds, angle–delay evolution in ASU drives larger short‑term errors than in the stochastic generators. Similar SH performances suggest that RT channels do not necessarily contain higher variability, but rather different. In fact, at longer horizons, a channel progression guided by angle-delay may even be easier to learn than purely Doppler-driven phase rotations.

\paragraph{\textbf{Cross-testing (5~ms horizon)}}
The top half of Table~\ref{tab:cross_finetune_stacked} shows zero‑shot transfer performance for a prediction horizon of 5~ms. Along the diagonal, we see in-domain NMSE ranking the targets by prediction difficulty as: CDL‑C ($-17.2$) $<$ TDL‑A ($-15.8$) $<$ UMa ($-12.2$) $<$ ASU ($-8.6$). Within the stochastic family, transfer is moderate but not symmetric, spanning $-11.5$ to $-15.0$~dB. Such strong zero-shot performances suggest the similarity of the task - understandable since in stochastic datasets, only Doppler progression must be learned. Noteably, crossing the stochastic–ray‑traced boundary fails in both directions: stochastic $\to$ASU yields positive NMSE ($11.0$, $12.2$, $7.8$~dB), and ASU$\to$stochastic is also positive ($1.9$, $2.7$, $4.0$~dB). This asymmetry is consistent with a representational mismatch—stochastic models lack spatial structure, so predictors trained there do not learn the angle–delay kinematics needed for ASU. A predictor trained on ASU does not directly map to the simpler stochastic dynamics, but it is closer. Overall, prediction at a long-enough horizon depends heavily on accurate modeling, with stochastic simplifications do not directly transfer to RT. Ray-tracing or GBSMs are thus necessary for realistic algorithm assessments.

\paragraph{\textbf{Fine-tuning (400 target samples; 5~ms horizon)}}
With 1\% of target data, fine‑tuning partially bridges the stochastic–ray‑traced gap: stochastic$\to$ASU‑1cm improves significantly, but absolute performances remain low (TDL‑A: $-3.2$, CDL‑C: $1.0$, UMa: $-3.4$~dB). ASU‑1cm$\to$stochastic shows better results: $-1.2$~dB in TDL‑A, $-4.1$~dB in CDL‑C, and $-1.2$~dB in UMa. By contrast, stochastic transfers with 1\% adaptation are mixed—implying small-sample fine-tuning can overfit or forget when source and target are already close. Among stochastic targets, CDL‑C remains the easiest and UMa the hardest at 5~ms; among sources, TDL‑A and UMa transfer to ASU more strongly than CDL-C. Contrary to the compression task, fine-tuning on more data is necessary to learn the angle-delay progression only present in RT.

\begin{figure}[t]
    \centerline{\includegraphics[width=1\columnwidth]{ch_pred_val_losses.png}}
    \caption{Channel prediction NMSE (dB) based on input sequence of 10 samples, one every ms, and single output at horizon $\Delta t\in\{1,3,5,10,20,40\}$~ms for three stochastic generators (TDL-A, CDL-C, UMa) and a ray-traced scene (ASU). Solid curves denote the best learned GRU-based predictor per dataset; dashed curves denote a sample-and-hold (S\&H) baseline. Lower means better channel prediction performance. }
    \label{fig:ch_pred}

    \vspace{-.4cm}
\end{figure}

\section{Key Takeaways \& Conclusion} \label{sec:takeaways_conclusion}

\noindent\textbf{Key takeaways} The bullets below condense answers to the three guiding questions in Section \ref{sec:intro}:
\begin{itemize}[leftmargin=*,nosep]
  \item \textbf{Q1—Sufficiency.} Training \& evaluating only on stochastic data yields \emph{unreliable} expectations for site‑specific behavior: depending on the profile, it can \emph{over‑ or under‑estimate} performance relative to RT (our proxy for deployment reality). Light adaptation (e.g., ~400 samples of the target RT scenario) can repair channel compression performance but is typically insufficient for prediction. The more limited the available data is, the more beneficial pre-training becomes, particularly when performed on geometrically similar RT, followed closely by diverse stochastic models (UMa/TDL).
  
  \item \textbf{Q2—Relative utility.} Stochastic datasets are useful for scalable pre-training and on tasks that have light requirements on channel structure. RT should be preferred (or GBSM with strong spatial consistency) when tasks depend on coupled delay–angle–Doppler dynamics. Transfer learning performance appears to follow dataset similarity: with RT$\rightarrow$RT transferability generally beating stochastic$\to$RT.
  
  \item \textbf{Q3—Model selection:} The suitability of a channel model for a given ML task remains an open question. In Section~\ref{sec:open-research}, Table~\ref{tab:task_model_core} offers an initial mapping from channel model assumptions to ML task requirements. We avoid definitive claims across all tasks and models: real-world models are missing, and even the two tasks analyzed here yield nuanced takeaways. That said, when ray-tracing is a reasonable proxy for the target environment, incorporating it into ML studies yields more credible results. Given the practicality of obtaining ray-traced data (e.g., DeepMIMO, SionnaRT), we encourage the community to validate their conclusions with it.
\end{itemize}

\paragraph*{\textbf{Conclusion}}
Across two representative tasks, we find that \emph{stochastic‑only} training and evaluation rarely resemble site‑specific behavior. For compression, good stochastic in‑domain scores do not imply strong performance on RT; for prediction, stochastic$\!\to$RT fails even at 1\,ms horizons. Ray‑traced data better captures the coupled delay–angle–Doppler evolution and spatial correlations that some wireless ML tasks rely on (like channel prediction). Nevertheless, stochastic datasets remain \emph{useful}: they provide scalable, reproducible pre-training (with UMa/TDL, not CDL) that adapts well to structure-light tasks (such as channel compression). The most reliable path to deployment, therefore, seems to be a \emph{hybrid} pipeline: use stochastic data for breadth and RT data for faithfulness, with evaluation/fine-tuning anchored on site-specific (and ultimately real) channels.

\paragraph*{\textbf{Future work}}
Two immediate steps are (i) \emph{validation on real‑world measurements} to quantify simulation‑to‑real gaps, and (ii) a \emph{larger‑scale RT study} spanning many scenes, mobility profiles, and materials to stress‑test transferability across tasks. A promising direction is a \emph{three‑level training}: (1) variability within the target deployment (multiple materials/assumptions for the same site), (2) variability across \emph{similar} deployments (near‑neighbors in a task‑aware similarity space), and (3) stochastic augmentation for additional diversity (following 3GPP guidelines on parameterizing stochastic models with ray-tracing large-scale parameters). This aims to harden models to both parametric uncertainty (materials, vegetation, building details, mobility patterns) and domain shift, while guiding the selection/creation of RT data that most closely resembles the intended deployment. Ultimately, we envision an evaluation standard where results are reported on (i) real data when available, (ii) site‑specific RT chosen via explicit similarity criteria, and (iii) selected stochastic baselines.

\begin{table}[t]
\centering
\caption{Channel Prediction NMSE (dB) on 5 ms horizon.}
\label{tab:cross_finetune_stacked}
\renewcommand{\arraystretch}{1.15}
\setlength{\tabcolsep}{4pt}

\begin{minipage}{\columnwidth}
\centering
\textbf{Cross-Test}\par\vspace{2pt}
\begin{tabular}{lcccc}
\toprule
\textbf{Training Model} & \textbf{TDL-A} & \textbf{CDL-C} & \textbf{UMa} & \textbf{ASU-1cm} \\
\midrule
TDL-A    & \textbf{-15.8} & -14.8 & -15.0 & 11.0 \\
CDL-C    & -14.9 & \textbf{-17.2} & -12.0 & 12.2 \\
UMa      & -12.1 & -11.5 & \textbf{-12.2} & 7.8 \\
ASU-1cm  & 1.9 & 2.7 & 4.0 & \textbf{-8.6} \\
\bottomrule
\end{tabular}
\end{minipage}

\vspace{4pt}

\begin{minipage}{\columnwidth}
\centering
\textbf{Fine-Tuning (1\%)}\par\vspace{2pt}
\begin{tabular}{lcccc}
\toprule
\textbf{Training Model} & \textbf{TDL-A} & \textbf{CDL-C} & \textbf{UMa} & \textbf{ASU-1cm} \\
\midrule
TDL-A    & \textbf{-15.8} & -13.0 & -12.7 & -3.2 \\
CDL-C    & -13.6 & \textbf{-17.2} & -10.3 & 1.0 \\
UMa      & -12.1 & -13.0 & \textbf{-12.2} & -3.4 \\
ASU-1cm  & -1.2 & -4.1 & -1.1 & \textbf{-8.6} \\
\bottomrule
\end{tabular}
\end{minipage}
\end{table}

\section{Open Research} \label{sec:open-research}

\begin{table*}[t]
\centering
\caption{Task–channel model suitability under typical configurations (\underline{theory-informed, indicative, not definite}) .
Symbols: \\ \cmark\ usable;
\mmark\ usable with caveats; \xmark\ not suitable.
Caveats C1–C6 in Sec.~II. \textbf{Evidence:} \textsf{ours} = studied in this paper; — = pending.}
\label{tab:task_model_core}
\renewcommand{\arraystretch}{1.1}
\setlength{\tabcolsep}{3.6pt}
\small
\begin{tabular}{p{4.0cm} p{6.9cm} c c c c c}
\toprule
\textbf{Task} & \textbf{Required channel properties} & \textbf{TDL} & \textbf{CDL} & \textbf{GBSM} & \textbf{RT} & \textbf{Evidence} \\
\midrule
Power/MCS selection & SNR, coarse fading stats & \cmark & \cmark & \cmark & \cmark & — \\
MIMO CSI compression & Intra-sample freq.\ coherence; array manifold (C1,C4) & \mmark & \cmark & \cmark & \cmark & \textsf{ours} \\
Antenna selection & Spatial correlation; per-user variation (C2) & \mmark & \cmark & \cmark & \cmark & — \\
Beam prediction & Angle structure; geometry continuity (C1,C2) & \xmark & \cmark & \cmark & \cmark & — \\
AoA/AoD estimation & Per-path angles; cluster structure (C1) & \xmark & \cmark & \cmark & \cmark & — \\
Localization & Delay/angle geometry; multi-link correlation (C1,C2) & \xmark & \mmark & \cmark & \cmark & — \\
Temporal prediction & Coupled delay–angle–Doppler evolution (C1,C2) & \xmark & \mmark & \cmark & \cmark & \textsf{ours} \\
Spatial prediction & Spatial continuity; visibility regions (C2,C3) & \xmark & \mmark & \cmark & \cmark & — \\
Handover prediction & Trajectory + multi-cell correlation (C2,C3) & \xmark & \xmark & \cmark & \cmark & — \\
Frequency translation & Cross-band structure; materials (C5) & \xmark & \xmark & \mmark & \cmark & — \\
Digital-twin training & Site specificity; materials; (C4,C5) & \xmark & \xmark & \mmark & \cmark & — \\
\bottomrule
\end{tabular}
\end{table*}

In this section, we move beyond the two tasks we studied empirically and sketch a broader, still-open mapping between ML tasks and channel models. 
Fig.~\ref{fig:determinism_continuum_full} shows how channel families move from amplitude/noise statistics to array- and geometry-aware models, along with the governing formulas. These equations make the channel assumptions of each model more explicit. To bridge these to ML task requirements, we review per-task modeling needs, starting from inputs and outputs, and then match those needs to the available structures (e.g., delay, angle, Doppler, spatial consistency, and materials):

\begin{itemize}[leftmargin=*]
  \item \textbf{Power/MCS selection} — a classifier/regressor from SNR/coarse CSI to an MCS or power level. \emph{Needs:} large-scale fading, basic SNR reliability. All model families suffice when the input excludes angle/geometry; adaptation logic is standardized in NR physical-layer procedures for data (CSI reporting)~\cite{3gpp38214_portal}. 

  \item \textbf{MIMO CSI compression} — learn a compact representation of $\mathbf H$ for feedback. \emph{Needs:} intra-sample frequency coherence and an array manifold; geometry helps but is not mandatory. Stochastic CDL, GBSM, and map-based data are all viable; realism and transfer improve with higher determinism~\cite{wen2017csinet}. 

  \item \textbf{Beam prediction} — map CSI (and/or side information) to a beam index/vector. \emph{Needs:} per-path angles and stable spatial correlation across time/links; this requires models with strong spatial consistency. Public map-based datasets (e.g., DeepMIMO) are widely used for this task~\cite{alkhateeb2019deepmimo}. 

  \item \textbf{AoA/AoD estimation} — infer per-path angles. \emph{Needs:} explicit angle structure and path clustering; GBSM/RT are natural fits; Cluster-based stochastic like UMa can be used with limitations~\cite{liu2012cost2100,jaeckel2014quadriga}.

  \item \textbf{Temporal channel prediction} — predict future $\mathbf H$ from a window of past samples. \emph{Needs:} coupled delay–angle–Doppler evolution and spatial consistency; TDLs / CDLs limit temporal change to Doppler within stationary clusters, whereas GBSM/RT preserve richer dynamics~\cite{gru_pred_mdpi}. 

  \item \textbf{Localization / spatial prediction / handover} — estimate position or future cell/beam from CSI and motion. \emph{Needs:} geometry-informed delay/angle, multi-link correlation, and possibly indoor–outdoor behavior. GBSM with spatial consistency or RT are preferred; RT is essential for material/penetration effects and accurate multipath modeling~\cite{alkhateeb2019deepmimo}.
\end{itemize}

\noindent Table~\ref{tab:task_model_core} aggregates these requirements and marks each channel model family as \emph{sufficient}, \emph{usable with caveats}, or \emph{unsuitable}. Here, "sufficient" refers to structures that the task fundamentally requires under standard settings; it does not preclude caveats or performance gaps in real-world deployments. Flags (e.g., spatial consistency, indoor–outdoor transitions, materials, near-field) indicate which properties must be enabled to make a model \emph{fit for the task}.


\bibliographystyle{IEEEtran}
\bibliography{biblio}

\end{document}

%% file: input.tex
\usepackage{amsfonts}
\usepackage{times}
\usepackage{graphicx}
\usepackage{latexsym}
\usepackage{dsfont}
\usepackage{amssymb}
\usepackage{amsmath}
\usepackage{cite}
\usepackage{verbatim}
\usepackage{subfigure}

\def\bb0{{\mathbb{0}}}

\def\bb{{\mathbf{b}}}

\def\b0{{\mathbf{0}}}

\def\sf0{{\mathsf{0}}}

\usepackage{epstopdf}
\usepackage{enumerate}
\usepackage{algorithmicx}
\usepackage{algorithm}
\usepackage{amsmath}
\usepackage[noend]{algpseudocode}
\usepackage{float}
\usepackage{hyperref}
\usepackage{color}
\usepackage{makeidx}
\usepackage{bbm}
\usepackage{graphicx}
\usepackage{booktabs}
\usepackage{subfigure}
\usepackage{multirow}

\usepackage{stackengine} 
\stackMath

\newlength\matfield
\newlength\tmplength

\usepackage{relsize}